\documentclass[12pt,reqno]{article}
\usepackage{amsfonts}
\usepackage[utf8x]{inputenc}
\usepackage[english]{babel}
\usepackage{amsmath,amssymb,amsfonts,wasysym}   
\usepackage{graphics}                           
\usepackage{amsmath}  
\usepackage{mathtools} 
\usepackage{hyperref}                           
\usepackage{comment}
\newtheorem{theorem}{Theorem}[section]

\newtheorem{corollary}[theorem]{Corollary}

\newtheorem{observation}[theorem]{Observation}

\newcommand*{\tr}{\mathop{}\!\mathrm{Tr}}
\newcommand*{\Proof}{\mathop{}\!\mathit{Proof}}

\usepackage{thmtools}
\usepackage{thm-restate}

\usepackage{hyperref}

\usepackage{cleveref}

\begin{document}
\title{Berezin quantization of Gaussian functions depending by a quantum and compression parameter}
\author{Simone Camosso$^*$}
\date{}
{\renewcommand{\thefootnote}{\fnsymbol{footnote}}
\setcounter{footnote}{1}
\footnotetext{\textbf{e-mail}: r.camosso@alice.it.\\ These notes on the Berezin quantization has been written 
when the author was a teacher at Liceo ``G.B.Bodoni'' (Cuneo, Piedmont, Italy).}
\setcounter{footnote}{0}
}

\maketitle

\begin{center}
\textbf{Abstract}
\end{center}
\noindent
The aim of this work is to study the Berezin quantization of a Gaussian 
state. The result is another Gaussian state that depends on a quantum parameter $\alpha$, 
that describes the relationship between the classical and quantum vision. The compression parameter 
$\lambda >0$ is associated to the harmonic oscillator semigroup. 

\vspace{1cm}


\section{Introduction}

This paper is devoted to the study of the quantization of Gaussian states. 
Let us consider a function $f(x,p)$, with $(x,p)\in \mathbb{R}^{2n}$, denoted by $B_{\alpha}(f)$ as its Berezin quantization. 
The choice of the Berezin quantization is due to the fact that we will consider Gaussian functions on $\mathbb{C}^{n}$ instead $\mathbb{R}^{2n}$ and, 
for this reason, a good scheme of quantization of $\mathbb{C}^{n}$ is the Berezin quantization. 
It is well known that this scheme of quantization in comparison with the Weyl quantization presents ``a few problems'': 
for example it doesn't preserve polynomial relations, the product rules 
are more complicated than Weyl quantization and the equivalent of Eherenfest 
theorem doesn't hold. Indeed this scheme of quantization is rarely used to describe the system dynamics. 
On the other hand, it leads naturally to the definition of a vacuum state $\psi_{0}$ that  is usually a Gaussian function. 

In the first two sections we will review basic notions on the Berezin quantization 
of $\mathbb{C}^{n}$ and the quantum harmonic oscillator, references are \cite{uno}, \cite{due}, \cite{tre}, \cite{cinque}, \cite{undici}, \cite{sei} and \cite{sette}. 
The other sections are devoted to the proofs of the following results.

\begin{theorem}
  \label{theo1}
In the setting of the Berezin quantization  of $\left(\mathbb{C}^n,\langle \cdot, \cdot \rangle\right)$ 
we have that the quantization of the complex Gaussian $f(z)\,=\,Ce^{-\frac{\lambda}{4} (z+\overline{z})^2}$, with $C\,=\,\frac{\sqrt{e}}{2^n}$, is given by:

\begin{equation}
 \label{berezinprop}
 B_{\alpha}(f)\,=\,C'\left(\frac{\alpha}{\alpha+\lambda}\right)^{\frac{n}{2}}e^{-\frac{\alpha\lambda}{4(\alpha+\lambda)}(z+\overline{z})^2},
\end{equation}
\noindent 
where $\alpha$ is a quantum parameter, $C'\,=\, \frac{\sqrt{e}}{2^n} $, $z\in\mathbb{C}^{n}$ and $\lambda>0$.
\end{theorem}

It would be desirable to evaluate the trace of the previous Berezin transform. Unfortunately following the definition of \cite{otto} 
for the trace of the Berezin transform we find an infinite value. 
As corollary we can consider instead the classical trace another kind of trace deriving from the usual inner product on the space  
$L^{2}(\Omega,\rho)$. In this case we have a ``modified'' version that is the trace of the squared Berezin symbol.

\begin{corollary}
  \label{cor1}
 Let $\widetilde{B_{\alpha}(f)}\,=\,\frac{B_{\alpha}(f)}{\sqrt{C'}\left(\frac{\alpha}{\alpha+\lambda}\right)^{\frac{n}{4}}}$, then we have that:
 
 \begin{equation}
   \tr\left(\widetilde{B_{\alpha}(f)}^2\right)\,=\, \left(\frac{\alpha}{\alpha+3\lambda}\right)^{\frac{n}{2}}.
 \end{equation}
\end{corollary}

We observe that the selected Gaussian function $f(z)$ depends by a ``compression 
factor'' $\lambda>0$ and that $\alpha$ is a quantum parameter that corresponds to the inverse of the Planck constant $h$. 
We observe that when $\alpha \rightarrow +\infty$ the trace tends to $1$. Moreover when $\alpha=\lambda=1$ we have $\frac{1}{2^{n}}$. This can be interpreted as an index of the purity of the state.
The next result I will present is inspired by the work of \cite{quattro} and \cite{dieci}, it corresponds to a ``generalized''  version of the Heisenberg principle in one dimension.

\begin{theorem}
 \label{heisenbergt}
 In the setting of the Berezin quantization of $\mathbb{C}$, if $\lambda >0$ then 
 
 $$ \sigma_{\psi_{\lambda}}^{2}(x)\sigma_{\psi_{\lambda}}^{2}(p)\,=\, \frac{1}{4}\left( -i[x,p]\psi_{\lambda},\psi_{\lambda}\right)^2,$$
 \noindent 
 where $\psi_{\lambda}\,=\, C'\left(\frac{\alpha}{\alpha+\lambda}\right)^{\frac{n}{2}}e^{-\frac{\alpha\lambda}{4(\alpha+\lambda)}(z+\overline{z})^2}$ 
 is the quantized Gaussian and $(\cdot , \cdot )$ the inner procuct of elements of $L^2(\mathbb{R})$.
\end{theorem}
 
In this theorem the quantum parameter has been fixed to $1$ as it is convention with the natural units. 

\section{The Berezin quantization}

Let $\Omega$ be a domain in $\mathbb{C}^n$ with the usual inner product $\langle z ,w \rangle\,=\, z\cdot \, \overline{w}\,=\, \sum_{j=1}^{n}z_{j}\overline{w_{j}} 
$. Let $\rho>0$ be a weight function on $\Omega$ and $L^{2}(\Omega,\rho)$ be the 
space of square integrable functions respect $\rho$. Let $L^{2}_{\text{hol}}(\Omega,\rho)$ 
be the subspace of square integrable holomorphic  functions respect to $\rho$. This 
space is also called the ``weighted Bergmann space'' and has a reproducing 
kernel $K_{\rho}(z,w)$. Let us assume $K_{\rho}(z,z)>0$ for all $z$, we define
the ``Berezin transform'' of $f\in L^{\infty}(\Omega)$ the following integral operator:

\begin{equation}
\label{berezin0}
B_{\rho}f(z)\,=\, 
\frac{1}{K_{\rho}(z,z)}\int_{\Omega}f(w)\left|K_{\rho}(z,w)\right|^2\rho(w)dwd\,\overline{w}.
\end{equation}

The Berezin transform is an important tool in the contest of Berezin 
quantization, especially its asymptotic behaviour with the appropriate weights 
$\rho$. As showed in \cite{due} the construction of the Berezin quantization 
reduces to constructing a family of weights for which the associated Berezin 
transform $B_{\rho}$ has an asymptotic expansion:

\begin{equation}
 \label{asymptotic}
 B_{\rho_{\alpha}}\,=\, Q_{0}+ \frac{1}{\alpha} Q_{1}+ \frac{1}{\alpha^{2}} Q_{2} + \ldots,
\end{equation}
\noindent 
where $\alpha\,=\, \frac{1}{h}$ is the `` formal parameter'' that when $h \rightarrow 0$ 
we have $\alpha \rightarrow +\infty$, $Q_{0}$ is the identity operator and $Q_{j}\,=\,\sum_{\beta,\gamma\in\mathbb{N}^n}c_{j\beta\gamma}\partial^{\beta}\overline{\partial}^{\,\gamma}(f)$
are differential operators with $\beta,\gamma$ multi--indices. From $Q_{j}$ it is possible to define the bidifferential operators 
$C_{j}(f,g)\,=\, \sum_{\beta,\gamma\in\mathbb{N}^n}c_{j\beta\gamma}\partial^{\beta}(f)\overline{\partial}^{\,\gamma}(g)$, 
and the star--product:

\begin{equation}
  \label{starprod}
  f\star g\,=\, \sum_{j=0}^{+\infty}\frac{1}{\alpha^{j}}C_{j}(f,g).
\end{equation}

If the condition $C_{1}(f,g)-C_{1}(g,f)\,=\, \frac{i}{2\pi}\{f,g\}$ it's valid then the star product coincides with the Berezin star--product  and this provides a Berezin 
quantization. Here $\{\cdot ,\cdot\}$ is the Poisson bracket on $\mathbb{C}^n$ and  $f,g\in C^{\infty}(\Omega)$ are quantum observables.

The proof of this assertion and many details on the Berezin quantization of $\mathbb{C}^{n}$ can be found in 
\cite{due}. For the Berezin quantization of general function spaces the reader can consult \cite{tre}. 

For our purpose we consider the Berezin quantization of $\mathbb{C}^{n}$ with the weighted Bergmann space as function space. 
In this case $\rho(z)\,=\, \left(\frac{\alpha}{\pi}\right)^{n}e^{-\alpha z\cdot \, 
\overline{z}}$, $\alpha$ will be the quantum parameter that tends to infinity and $K_{\rho}(z,w)\,=\, e^{\alpha z\cdot \, \overline{w}}$. 

\section{The quantum harmonic oscillator}

We consider a slightly modified version of the quantum harmonic oscillator in $\mathbb{C}^{n}$. Let 
$H\,=\, \widehat{x}^2+\widehat{p}^2 \,=\, \sum_{j=1}^{n} \widehat{x}_{j}^2+ \widehat{p}_{j}^2$ be the Hamiltonian operator where 
$\widehat{x}_{j}\,=\, x_{j}$ and $\widehat{p}_{j}\,=\, -i\partial_{x_{j}}$ are the usual quantum operators that satisfy the following commutation relations:

\begin{equation}
  \label{canonicalrelations}
 [x_{j},p_{k}]\,=\,i h\delta_{jk}, [x_{j},x_{k}]\,=\,  [p_{j},p_{k}]\,=\,0,
 \end{equation} 
\noindent 
for every $j,k\,=\, 1, \ldots , n$ and where $h$ is the Plank constant. We 
define the operators:

\begin{equation}
  \label{canonicalrelations}
 \widehat{z}_{j}\,=\, \frac{\widehat{x}_{j}+i\widehat{p}_{j}}{\sqrt{2}}, \ \  \widehat{\overline{z}}_{j}\,=\, \frac{\widehat{x}_{j}-i\widehat{p}_{j}}{\sqrt{2}},
 \end{equation} 
\noindent 
for every $j=1, \ldots , n$. The operators $\widehat{z},\widehat{\overline{z}}$ 
are called respectively the annihilation and creation operator. In this notation the Hamiltonian assume the following form $H\,=\, \sum_{k=1}^{n}2\widehat{\overline{z}}_{k}\widehat{z}_{k}+h$. 
It is possible to prove that the ground state corresponding to the energy level $E_{0}=nh$ is given by 
the Gaussian:

\begin{equation}
  \label{groundstate}
  \psi_{0}\,=\, e^{-\frac{x^2}{2}},
\end{equation}
\noindent 
where $x^2\,=\, \sum_{k=1}^{n} x_{k}^{2}$.
In general we have eigenvalues of energy in this even form:

\begin{equation}
  \label{eigenvalues}
  E_{j}\,=\, n\cdot (2j+h),
\end{equation}
\noindent 
with eigenfunctions given by $\psi_{j}\,=\, \prod_{k=1}^{n}x_{k}^{j}e^{-\frac{x^2}{2}}$.

\section{Proof of the theorem $\ref{theo1}$}

$\Proof.$ The Berezin transform of $e^{-\frac{\lambda}{4} \left(z+\overline{z}\right)^{2}}$ with parameter $\alpha$ 
is:

\begin{equation}
  \label{berezinproof1}
B_{\rho}(f)\,=\, \left(\frac{\alpha}{\pi}\right)^{n}e^{-\alpha\langle w,w \rangle}\int_{\mathbb{C}^n}e^{-\frac{\lambda}{4}\left(z+\overline{z}\right)^{2}}\left|e^{\alpha\langle w,z\rangle}\right|^2e^{-\alpha\langle 
z,z\rangle}\,dzd\,\overline{z}.
\end{equation}

We remember the definition of the complex inner product $\langle w , z \rangle \,=\, w\cdot \, \overline{z} \,=\,\sum_{j=1}^{n}w_{j}\,\overline{z_{j}}$ 
and, after an algebraic semplification, we have that:

\begin{equation}
  \label{berezinproof2}
B_{\rho}(f)\,=\, \left(\frac{\alpha}{\pi}\right)^{n}e^{-\alpha w\cdot \,\overline{w}}\int_{\mathbb{C}^{n}}e^{-\frac{\lambda}{4}\left(z+\overline{z}\right)^{2}+\alpha w\,\overline{z}+\alpha \overline{w}z-\alpha z\,\overline{z}}\,dzd\,\overline{z}.
\end{equation}

This is a complex--Gauss integral depending by a quantum parameter $\alpha$ and the positive parameter $\lambda$. A simple way to solve the integral $(\ref{berezinproof2})$ consists to transform the complex integral in a real integral using the identification 
$\mathbb{C}^{n}\,\equiv\,\mathbb{R}^{2n}$. This gives the product of two real Gauss integrals:

\begin{equation}
  \label{berezinproof3}
  \begin{multlined}[t][12.5cm]
   B_{\rho}(f)\,=\, \left(\frac{\alpha}{\pi}\right)^{n}e^{-\frac{\alpha}{2} \left(x'^2+p'^2+\frac{1}{\alpha}\right)-\frac{1}{2}}\int_{\mathbb{R}^{n}}e^{-\left(\sqrt{\frac{\lambda+\alpha}{2}}x-\frac{\alpha}{2\sqrt{\frac{\lambda+\alpha}{2}}}x'\right)^2+ \frac{\alpha^2}{2(\lambda+\alpha)}x'^2}dx \\
   \cdot \int_{\mathbb{R}^n}e^{-\frac{\alpha}{2}(p^2-2pp'+p'^2)+\frac{\alpha}{2}p'^2}dp.
\end{multlined}
\end{equation}
\noindent 
where $z\,=\, \frac{x+ip}{\sqrt{2}}$, $\overline{z}\,=\, \frac{x-ip}{\sqrt{2}}$,  $w\,=\, \frac{x'+ip'}{\sqrt{2}}$, $\overline{w}\,=\, \frac{x'-ip'}{\sqrt{2}}$and 
$[x,x']=[p,p']=0$, $[x,p]=[x',p']=ih$ are the canonical relations. Adjusting the 
exponents of the two integrals we get two Gauss integrals that can be 
evaluated:

\begin{equation}
  \label{berezinproof4}
  \begin{multlined}[t][12.5cm]
   B_{\rho}(f)\,=\, \left(\frac{\alpha}{\pi}\right)^{n}e^{-\frac{\alpha}{2} \left(x'^2+p'^2-\frac{1}{\alpha}\right)-\frac{1}{2}+ \frac{\alpha^2}{2(\lambda+\alpha)}x'^2+\frac{\alpha}{2}p'^2}\cdot \frac{\pi^{\frac{n}{2}}2^{\frac{n}{2}}}{(\lambda+\alpha)^{\frac{n}{2}}}
   \cdot \frac{\pi^{\frac{n}{2}}2^{\frac{n}{2}}}{\alpha^{\frac{n}{2}}}.
\end{multlined}
\end{equation}
Thus we find that the initial integral is equal to $2^{n}\left(\frac{\alpha}{\alpha+\lambda}\right)^{\frac{n}{2}}e^{-\frac{\alpha \lambda}{2(\alpha+\lambda)}x'^2-1}$.

\hfill $\Box$

\begin{observation}
We observe that when $\alpha \rightarrow +\infty$ the complex quantized Gaussian 
$C'\left(\frac{\alpha}{\alpha+\lambda}\right)^{\frac{n}{2}} e^{-\frac{\alpha\lambda}{4(\alpha+\lambda)}(z+\overline{z})^2}$ tends to the classical complex Gaussian $C'e^{-\frac{\lambda}{4}(z+\overline{z})^2}$. 
\end{observation}

\begin{observation}
We can rewrite the $\left(\frac{\alpha}{\alpha+\lambda}\right)^{\frac{n}{2}} e^{-\frac{\alpha\lambda}{4(\alpha+\lambda)}(z+\overline{z})^2}$ as

$$ \left[\frac{1}{\left(1+\frac{\lambda}{\alpha}\right)^{\frac{n}{2}}}e^{-\frac{\lambda^2}{4\alpha\left(1+\frac{\lambda}{\alpha}\right)}(z+\overline{z})^2}\right] e^{-\frac{\lambda}{4}(z+\overline{z})^2}$$
\noindent 
and Taylor expand the square brackets:

$$ \left[1+\frac{\lambda^2}{4\alpha}(z+\overline{z})^2-\frac{n}{2}\cdot\frac{\lambda}{\alpha}+ \ldots \right] e^{-\frac{\lambda}{4}(z+\overline{z})^2}.$$

If we not consider the term $\frac{\sqrt{e}}{2^n}$, this is exactly the heat solution operator $Be^{-\frac{\lambda}{4}(z+\overline{z})^2}\,=\, e^{\frac{\Delta}{4\alpha}}e^{-\frac{\lambda}{4}(z+\overline{z})^2}$ 
according to \cite{due}, where $\Delta$ is the complex Laplacian on $\mathbb{C}^{n}$ 
given by $\Delta\,=\, 
4\sum_{j=1}^{n}\partial_{w_{j}}\partial_{\overline{w_{j}}}$.
\end{observation}

\section{Proof of the Corollary $\ref{cor1}$}

By the modified version of the trace (``that is a sort of a trace of a square'') 
we have that:

\begin{equation}
  \tr(B_{\alpha}(f)^2)\,=\,\left(\frac{\alpha}{\pi}\right)^{n}\int_{\mathbb{C}^n}B_{\alpha}(f)(z)^2e^{-\alpha z\cdot 
  \,\overline{z}}dzd\,\overline{z}.
\end{equation}

Thus we must to evaluate the integral:

\begin{equation}
  \tr(B_{\alpha}(f)^2)\,=\,\left(\frac{\alpha}{\pi}\right)^{n}\int_{\mathbb{C}^n}C'^2\left(\frac{\alpha}{\alpha+\lambda}\right)^{n}e^{-\frac{\alpha \lambda}{2(\alpha+\lambda)}(z+\overline{z})^2} e^{-\alpha z\cdot 
  \,\overline{z}}dzd\,\overline{z}.
\end{equation}

Using the canonical relations, this can be written as:

\begin{equation}
\begin{multlined}[t][12.5cm]
 \tr(B_{\alpha}(f)^2)\,=\,\frac{e}{2^{2n}}\left(\frac{\alpha}{\pi}\right)^{n}\left(\frac{\alpha}{\alpha+\lambda}\right)^{n}\int_{\mathbb{R}^{n}}e^{\frac{-3\alpha \lambda 
  -\alpha^2}{2(\alpha+\lambda)}x^2}dx\cdot 
  \int_{\mathbb{R}^n}e^{-\frac{\alpha}{2}p^2-\frac{1}{2}}dp.
    \end{multlined}
  \end{equation}

After the evaluation the Gauss integrals we find that the initial trace is equal to $C'\left(\frac{\alpha}{\alpha+\lambda}\right)^{\frac{n}{2}}\cdot 
\left(\frac{\alpha}{\alpha+3\lambda}\right)^{\frac{n}{2}}$.
Defining $\widetilde{B_{\alpha}(f)}\,=\,\frac{B_{\alpha}(f)}{\sqrt{C'}\left(\frac{\alpha}{\alpha+\lambda}\right)^{\frac{n}{4}}}$ and repeating calculations we find the result.
  \hfill $\Box$

\section{Proof of theorem $\ref{heisenbergt}$}
Before to start let us to fix some notation. We consider a generic Gaussian 
state with $\psi_{\lambda}\,=\, Ke^{-\frac{\lambda}{4(1+\lambda)}(z+\overline{z})^2}$ 
where $K\,=\,\frac{1}{\sqrt{e}}\left(\frac{\alpha}{\alpha+\lambda}\right)^{\frac{n}{2}}$ in dimension $n=1$. 
We denote by $\sigma_{\psi_{\lambda}}^{2}(x),\sigma_{\psi_{\lambda}}^{2}(p)$ 
respectively the variance of the observable $x$ and $p$. We have that:

\begin{equation}
 \label{heisenbergnew1}
 \sigma_{\psi_{\lambda}}^{2}(x)\,=\, 
 \int_{-\infty}^{+\infty}x^2K^2e^{-\frac{\lambda}{1+\lambda}x^2}dx\,=\, 
 \frac{K^2\sqrt{\pi}(1+\lambda)^{\frac{3}{2}}}{2\lambda^{\frac{3}{2}}},
 \end{equation}
\noindent 
where for the evalutation it is usefull the general result in one dimension:

$$ \int_{-\infty}^{+\infty}x^ne^{-ax^2}dx\,=\, \frac{1\cdot 3\cdot 5 \cdots 
(n-1)\sqrt{\pi}}{2^{\frac{n}{2}}a^{\frac{n+1}{2}}}, \ \ \ \ n\,=\, 2,4,6,\ldots 
, a>0.$$

Now we must evaluate $\sigma_{\psi_{\lambda}}^{2}(p)$. In a similar way as before 
we have:

\begin{equation}
 \label{heisenbergnew2}
 \begin{multlined}[t][12.5cm] \sigma_{\psi_{\lambda}}^{2}(p)\,=\, 
 \int_{-\infty}^{+\infty}K^2\left(\partial_{x}e^{-\frac{\lambda}{2(1+\lambda)}x^2}\right)^2dx\\ \,=\, 
 K^2\frac{\lambda^2}{(1+\lambda)^2}\int_{-\infty}^{\infty}x^2e^{-\frac{\lambda}{(1+\lambda)}x^2}
 \,=\,\frac{K^2\lambda^2\sqrt{\pi}(1+\lambda)^{\frac{3}{2}}}{2(1+\lambda)^2\lambda^{\frac{3}{2}}}.
 \end{multlined}
 \end{equation}

In order to prove the Heisenberg principle we must evaluate:

\begin{equation}
 \label{heisenbergnew3}
 \frac{1}{4}\left(K^2\int_{-\infty}^{+\infty}e^{-\frac{\lambda}{1+\lambda}x^2}dx\right)^2\,=\, \frac{K^4\pi (1+\lambda)}{4\lambda}.
 \end{equation}

Now by the ordinary Heisenberg principle:

$$\sigma_{\psi_{\lambda}}^{2}(x)\cdot \sigma_{\psi_{\lambda}}^{2}(p)\geq \frac{1}{4}\left(K^2\int_{-\infty}^{+\infty}e^{-\frac{\lambda}{1+\lambda}x^2}dx\right)^2,$$
\noindent 
and substituting our quantities we find that:

$$ \frac{1+\lambda}{\lambda}\geq  \frac{1+\lambda}{\lambda},$$
\noindent 
this is true for every compression $\lambda >0$.
\hfill $\Box$

\end{document}